\documentstyle [12pt] {article}

\begin{document}

\baselineskip = 20 true pt

\title{\huge \bf Dynamic critical properties of a one-dimensional probabilistic
cellular automaton}

\author{~\\ ~\\ {\LARGE Pratip Bhattacharyya}
                        \thanks{E-mail : pratip@hp1.saha.ernet.in}\\ ~\\
 \large Low Temperature Physics Section\\
         Saha Institute of Nuclear Physics\\
         Sector - 1, Block - AF, Bidhannagar\\
         Calcutta 700 064, India.}

\date{}

\maketitle

\vspace{1.5cm}

\begin{abstract}

\noindent Dynamic properties of a one-dimensional probabilistic cellular
automaton are studied by monte-carlo simulation near a critical point
which marks a second-order phase transition from a active state to
a effectively unique absorbing state. Values obtained for the
dynamic critical exponents indicate that the transition belongs to
the universality class of directed percolation. Finally the model
is compared with a previously studied one to show that a difference
in the nature of the absorbing states places them in different
universality classes.

\end{abstract}

\newpage

\section{ Introduction}

\indent Discrete models of nonequilibrium stochastic processes
form a class of interacting particle systems~\cite{Ligget1985}.
Of the models studied with short range and translationally invariant
interactions in space and time, the ones exhibiting a continuous
phase transition from an active steady state to an absorbing state
fall into two universality classes~\cite{Grassberger1995}~:

\noindent (1) the class of directed percolation (DP),\\
\noindent (2) the class of parity conservation (PC).

\indent Models with a unique absorbing state have been conjectured to
belong to the DP class~\cite{Janssen1981, Grassberger1982}. This is
yet the larger of the two classes and includes, for example, lattice
models of directed percolation in $d + 1$ dimensions~\cite{Kinzel1983},
the contact process for an epidemic~\cite{Harris1974}, Schl\"{o}gl's
first and second models of autocatalytic reactions~\cite{Schlogl1972},
the Domany-Kinzel automaton~\cite{Domany1984, Kinzel1985}, a lattice
version of reggeon field theory~\cite{Grassberger1979} and branching
annihilating random walks with an odd number of offsprings
\cite{Jensen1993-1}. The order parameter is usually the density of
\lq particles\rq~(occupied sites, kinks). The models mentioned above are
one-component systems and therefore the order parameter is scalar,
a requirement of the DP-conjecture~\cite{Janssen1981, Grassberger1982}.
The DP-conjecture was generalised to include multicomponent
systems such as the ZGB model of heterogeneous catalysis
\cite{Ziff1986, Grinstein1989, Jensen1990} and interacting
dimer-trimer models~\cite{Albano1995}. The class of DP was shown to
further include systems with an infinite number of absorbing states
\cite{Jensen1993-2}; these absorbing states are frozen configurations
characterised by unique statistical properties, a lack of long range
correlations and a general lack of symmetry among them.

\indent On the other hand, models with degenerate, mutually symmetric
absorbing states are believed to belong to the parity-conserving class
where the number of \lq particles\rq~are conserved modulo 2. Prominent
examples are the probabilistic cellular automaton models $A$ and $B$
of~\cite{Grassberger1984, Grassberger1989-1}, an interacting monomer-
dimer model~\cite{Kim1994} and branching annihilating random walks (BAW)
with an even number of offsprings~\cite{Jensen1994}. According to
one point of view~\cite{Grassberger1995, Grassberger1989-1} the mechanism
that puts these models in a class different from that of DP is the
conservation of particle number modulo 2. This point was proved for BAW
with an even number of offsprings: introduction of spontaneous
annihilation of particles in the model destroyed the conservation of
their number modulo 2 and the critical behaviour of this modified model
was observed to be in the class of DP~\cite{Jensen1993-3}.
According to a second point of view the critical behaviour of the models
in the PC class is due to the symmetry among its absorbing states
~\cite{Park1995, Hinrichsen1997}. To prove the point the interacting
monomer-dimer model of~\cite{Kim1994}, in the presence of a weak
parity-conserving field that destroyed the symmetry among the absorbing
states, was shown to exhibit critical behaviour in the DP class
~\cite{Park1995}. This view was further emphasized in the generalised
versions of the Domany-Kinzel automaton and the contact process
~\cite{Hinrichsen1997}. In these generalised models there was no explicit
parity-conservation law and with two symmetric absorbing states the
critical behaviour was in the PC class. In the presence of a symmetry
breaking field the critical behaviour of these models changed to the
DP class. However in the case of the probabilistic cellular automata
of~\cite{Grassberger1984, Grassberger1989-1} both the conservation
of particle number modulo 2 and mutual symmetry among the absorbing states
are present and therefore it it is not apparent which of these two features
is responsible for the models to be in the PC class.

\indent In this paper I shall study, using time-dependent monte-carlo
simulations, the dynamic critical properties of a one-dimensional
probabilistic cellular automaton~\cite{Bhattacharyya1996} which has
three absorbing states and exhibits a phase transition from a active
state to one of them only. The simulations provide values for the critical
point (more accurate than previous estimates~\cite{Bhattacharyya1996,
Bhattacharyya1997}) and the dynamic critical exponents that decide
the universality class to which the phase transition belongs.

\section{ The model}

\indent The probabilistic cellular automaton studied here is
\lq elementary\rq~(in the sense of Wolfram~\cite{Wolfram1983})
with two states per site and translationally invariant nearest neighbour
interactions. The probabilistic behaviour enters the model through
two mutually symmetric components of the evolution rule (like specific
noise added to a otherwise deterministic system), the rest of the rule
components being deterministic in nature.

\indent Formally the model~\cite{Bhattacharyya1996} is defined as a
line of sites with a binary variable $x_i \in \{0, 1\}$ assigned to
each site $i$. A site is said to be occupied if $x_i = 1$ and
unoccupied otherwise. Starting from a given configuration
$\{x^{(0)}_i\}$ the system evolves by parallel update of the variable
$x_i$ at all lattice sites following a local rule of evolution.
With nearest neighbour interactions the evolution rule is
defined by a set of eight components $[x^{(t)}_{i-1},~x^{(t)}_i,
~x^{(t)}_{i+1}] \mapsto x^{(t+1)}_i$ corresponding to $2^3$ distinct
three-site neighbourhoods :

\begin{equation}
{t : \over {t+1 :}}~~~~~~{111 \over 0}~~~~{110 \over 0}~~~~{011 \over 0}~~~~{101 \over 0}~~~~{010 \over 1}~~\underbrace{{100 \over  }~~~~{001 \over  }}_{\begin{array}{lll}
                                                                                                                                                                  1 & \mbox{with probability} & p\\
                                                                                                                                                                  0 & \mbox{with probability} & 1 - p
                                                                                                                                                           \end{array}}~~{000 \over 0}
\label{eq:model}
\end{equation}

\vspace{0.5cm}

\noindent The evolution rule (\ref{eq:model}) thus follows, according to
Wolfram's nomenclature scheme~\cite{Wolfram1983}, \lq elementary\rq
~rule $4$ with probability $1 - p$ and \lq elementary\rq~rule $22$ with
probability $p$.

\indent Clearly, the components of the evolution rule are of two kinds:
$(1)$ active components, where the central site changes its value, and
$(2)$ passive components, where the value of the central site remains
unchanged. The dynamic evolution of the system is due to the active
components. This involves two opposing processes :

\noindent $(a)$ Annihilation of adjacent occupied sites
(multi-\lq particle\rq~annihilation) due to the rule components
$111 \mapsto 0,~110 \mapsto 0,~011 \mapsto 0$ prevents the survival
of occupied pairs of nearest neighbours; this is a deterministic process.

\noindent $(b)$ Creation of an occupied site $(100 \mapsto 1,~001 \mapsto 1)$
requires an unoccupied site to have exactly one occupied neighbour; this
process of creation occurs only with a probability $p$.
The multi-\lq particle\rq~annihilation and creation of \lq particles\rq~
can, in effect, give rise to a diffusion process $-$ if a
\lq particle\rq~at site $i$ gets annihilated after creating
another \lq particle\rq~at a neighbouring site $i + r$, it has
effectively taken a step from $i$ to $i + r$.

\indent The passive components determine the absorbing states of the model.
Since there is no spontaneous creation of occupied sites $(000 \mapsto 0)$
the \lq vacuum\rq~(all sites unoccupied) is always an absorbing
state :

\begin{equation}
\makebox[5cm][l]{Absorbing state I :}
x_i = \left . \begin{array}{ll}
               0 & \mbox{for all $i$.}
               \end{array}
      \right .
 \label{eq:absorb1}
\end{equation}

\noindent Again, it is evident from the rule components $010 \mapsto 1$
and $101 \mapsto 0$ that an occupied site with unoccupied neighbours
remains occupied and vice versa. These features of the evolution rule
lead to two mutually symmetric absorbing states :

\begin{equation}
\makebox[5cm][l]{Absorbing State II :}
x_i = \left\{ \begin{array}{ll}
                  0 & \mbox{for $i$ = even,} \\
                  1 & \mbox{for $i$ = odd.}
                  \end{array}
         \right .
 \label{eq:absorb2}
\end{equation}

\begin{equation}
\makebox[5cm][l]{Absorbing State III :}
x_i = \left\{ \begin{array}{ll}
                  1 & \mbox{for $i$ = even,} \\
                  0 & \mbox{for $i$ = odd.}
                  \end{array}
      \right .
 \label{eq:absorb3}
\end{equation}

\indent The main point of interest in this paper is a phase
transition exhibited by the model~\cite{Bhattacharyya1996}.
For $p$ less than a critical value $p_c$ there exists three
distinct steady states given by the three absorbing states of
the model. For all initial states but two, it has been observed in
computer simulations~\cite{Bhattacharyya1996} that the only steady
state is the \lq vacuum\rq~(absorbing state I). The two cases of exception
occur when the initial state is either absorbing state II or absorbing
state III, which must also be respectively the steady states of the
system. Evolving from any other initial state the mutually symmetric
absorbing states II and III are never reached. This can be understood
by the fact that these states are fixed points that repel
 $-$ a damage introduced in these two states by flipping
only a single bit spreads through the entire lattice and eventually the
\lq vacuum\rq~is reached. Above $p_c$ there is another steady state
called the \lq active\rq~state with a constant non-zero density of
occupied sites. The active state is stable only on an infinite lattice;
for finite lattices it occurs as a metastable state that will decay
to the \lq vacuum\rq~if allowed to evole for sufficiently long time.
On an infinite lattice, in the supercritical region $(p > p_c)$,
all possible initial states other than the three absorbing states evolve
into the \lq active\rq~steady state. The density of occupied sites
$\rho_\infty$ in the steady state acts as the order parameter for this phase
transition $-$ in the supercritical region $\rho_\infty$ goes to zero
continuously as $p$ approaches $p_c$~\cite{Bhattacharyya1996} :

\begin{equation}
\rho_\infty \sim (p - p_c)^{\beta},~~~~~~~p \rightarrow p_{c \, +} ,
\label{eq:order}
\end{equation}

\noindent where $\beta$ is the critical exponent for the order parameter.
For random initial states the model therefore makes a continuous
(second order) phase transition, at $p = p_c$, from a active state
with $\rho_\infty > 0$ to a effectively unique absorbing state, the
\lq vacuum\rq. Because of the conjecture of~\cite{Janssen1981,
Grassberger1982} this phase transition is expected to belong to
the universality class of DP.

\section{Dynamic properties at the critical point}

\indent The phase transition is characterised here by critical exponents
describing the dynamic properties of the model at the point of
transition (the critical point $p_c$). To do so dynamic properties
of the model are studied by monte-carlo simulations on a computer,
only close to $p_c$. While the study of steady state properties
require simulations starting from disordered states,
the dynamic properties are studied using initial states with a single
occupied site. Following the evolution rule (\ref{eq:model}) an
initial occupied site grows into a cluster; the position of this
initial occupied site is called the \lq origin\rq~of the cluster.
For each value of $p$, $10^4$ clusters were simulated. Each cluster
was allowed to evolve for $5000$ time steps, unless it had died out
earlier. Typical examples of evolution near the critical point are
shown in figure 1.

\indent The quantities measured are : $(1)$ the survival probability
$P(t)$, which gives the chance that there is at least one occupied site
after $t$ time steps, $(2)$ the average number of occupied sites $N(t)$
after $t$ time steps, and $(3)$ the mean square radius $R^2 (t)$
of the cluster (or, the mean square displacement from the \lq origin\rq~
of the cluster) after $t$ time steps. At the critical point $p = p_c$,
these quantities are expected to follow power-type scaling laws in the
long-time limit $(t \rightarrow \infty)$ :

\begin{eqnarray}
P(t) &\sim& t^{- \delta}, \nonumber \\
N(t) &\sim& t^{\eta}, \label{eq:dynamic} \\
R^2 (t) &\sim& t^z, \nonumber
\end{eqnarray}

\noindent where $\delta$, $\eta$ and $z$ are dynamic critical exponents.
In the case of $N(t)$ the average is taken over all clusters including
those which have died out, while $R^2 (t)$ is averaged over the occupied
sites in the surviving clusters only.

\indent Results for the three quantities $P(t)$, $N(t)$ and $R^2 (t)$,
obtained from computer simulations of the model (\ref{eq:model}), are
shown in figure 2. On log-log plot curves in the subcritical region
bend downward while those in the supercritical region bend upward;
at the critical point the curves are expected, if the scaling laws
(\ref{eq:dynamic}) are true, to be straight lines as $t \rightarrow \infty$.
It is obvious from figure 2 that $0.75 < p_c < 0.753$.
More precise estimates for $p_c$ and the critical exponents are obtained
by the method of effective exponents due to Grassberger
~\cite{Grassberger1989-2}. Effective exponents $\delta_t$, $\eta_t$
and $z_t$ are defined as the local slopes of the curves shown in figure 2 :

\begin{equation}
- \, \delta_t = {\Delta [\log P(t)] \over \Delta [\log t]}~,~~~~~~
\eta_t = {\Delta [\log N(t)] \over \Delta [\log t]}~,~~~~~~
z_t = {\Delta [\log R^2(t)] \over \Delta [\log t]}~,
\label{eq:effectdef}
\end{equation}

\noindent which are measured by using the formulae~\cite{Grassberger1989-2} :

\begin{equation}
- \, \delta_t = {\log[P(t) / P(t/m)] \over \log m}~,    \label{eq:effective1}
\end{equation}

\noindent and similar expressions for $\eta_t$ and $z_t$. Results for
$m = 10$ are shown in figure 3. Like the curves in figure 2, curves for
$p < p_c$ bend downward while those for $p > p_c$ bend upward.

\indent For finite times the dynamic quantities do not have a pure
power-law scaling form like (\ref{eq:dynamic}); in general there exists
correction-to-scaling of the type :

\begin{equation}
P(t) \sim t^{- \delta} \left(1 + {a \over t} + {b \over t^{\delta '}} + \cdots \right),
\label{eq:correction}
\end{equation}

\noindent and similar expressions for $N(t)$ and $R^2 (t)$ with
correction-to-scaling exponents $\delta '$, $\eta '$ and $z '$
respectively. Consequently the behaviour of the effective exponents
defined in (\ref{eq:effectdef}) are given by~\cite{Grassberger1989-2} :

\begin{equation}
\delta_t = \delta + {a \over t} + {\delta ' b \over t^{\delta '}} + \cdots ,
\label{eq:effective2}
\end{equation}

\noindent and similar expressions for $\eta_t$ and $z_t$. The critical
exponents of the model appear as the asymptotic values of the corresponding
effective exponents :

\begin{equation}
 \begin{array}{lll}
               \makebox[3.2cm][l]{$\delta = \lim_{t \rightarrow \infty} \delta_t$~,}
             & \makebox[3.2cm][l]{$\eta = \lim_{t \rightarrow \infty} \eta_t$~,}
             & \makebox[3.2cm][l]{$z = \lim_{t \rightarrow \infty} z_t$~.}
 \end{array}
\label{eq:asymptote}
\end{equation}

\noindent Therefore in a plot of an effective exponent (as the ordinate)
versus $1/t$ (as the abscissa) the corresponding critical exponent
is obtained as the intercept of the curve for $p = p_c$ on the
ordinate axis. Using this method the following estimates for the
critical characteristics of the model were obtained from computer
simulation data :

\begin{equation}
p_c = 0.7515 \pm 0.0005    \label{eq:criticalpoint}
\end{equation}

\noindent and

\begin{eqnarray}
\delta &=& 0.16 \pm 0.01, \nonumber \\
\eta &=& 0.32 \pm 0.02, \label{eq:criticalvalue} \\
z &=& 1.27 \pm 0.01. \nonumber
\end{eqnarray}

\noindent The value of the critical point agrees closely with
previous estimates~\cite{Bhattacharyya1996, Bhattacharyya1997}
and improves upon them. The values of the critical exponents are
also found to satisfy the scaling relation~\cite{Grassberger1979} :

\begin{equation}
d \: z = 2 \eta + 4 \delta,    \label{eq:scale1}
\end{equation}

\noindent where $d$ is the number of spatial dimensions of the
system (here $d = 1$).

\indent Evolving from disordered initial states, the density of
occupied sites $\rho_t$ was observed to follow the same dynamic
scaling as the survival probability. In computer simulations of
\cite{Bhattacharyya1996} the exponent characterising the critically
slow relaxation $\rho_t \sim t^{- \alpha}$ at $p = p_c$ was found to
be $\alpha \approx 0.16 \approx \delta$.

\indent The values of all the three dynamic critical exponents
agree, within the limits of error, with the corresponding values
for DP in $1 + 1$ dimensions~\cite{Grassberger1979, Essam1988,
Dickman1991}. In a previous work~\cite{Bhattacharyya1997}
the critical exponents $\nu_\perp$ and $\nu_\parallel$ for the
correlation length and correlation time respectively at this particular
phase transition, obtained by finite-size scaling methods,
were also found to belong to the DP class. However, the order parameter
exponent $\beta$ was observed to disagree with the DP value
\cite{Bhattacharyya1996}; this was an error arising out of
finite-size effects and fluctuations due to the small sample
size used for averaging. Since $\delta$ and $\nu_\parallel$
are already in the class of DP, the exponent $\beta$, by virtue
of the relation $\beta = \delta \, \nu_\parallel$
\cite{Grassberger1979}, must also agree with the directed
percolation value for $1 + 1$ dimensions.

\indent The last result is concerned with the fractal dimension
of the clusters in the single space dimension at $p = p_c$.
The average number of occupied sites per surviving cluster is
given by $N_s(t) = N(t) \, / \, P(t)$. The fractal dimension
$d_F$ of the clusters at fixed time is defined by :

\begin{equation}
N_s \sim R \, ^{d_F}.    \label{eq:fractal}
\end{equation}

\noindent Following the definition of the dynamic exponents
(\ref{eq:dynamic}) the fractal dimension is expected to satisfy
the relation :

\begin{equation}
d_F \: z = 2(\eta + \delta).    \label{eq:scale2}
\end{equation}

\noindent Figure 4 shows a log-log plot of $N_s$ versus $R$ at $p = p_c$.
The slope of the curve is given by :

\begin{equation}
d_F = 0.74 \pm 0.02,
\end{equation}

\noindent which satisfies relation (\ref{eq:scale2}) within the limits
of error.

\section{Discussion}

\indent In this concluding section I shall compare the model 
defined by (\ref{eq:model}) with another that has the same set
of absorbing states and that belongs to a different university
class.

\indent The probabilistic cellular automaton (\ref{eq:model})
studied here was found to have three absorbing states given
by (\ref{eq:absorb1}), (\ref{eq:absorb2}) and (\ref{eq:absorb3}).
For $p < p_c$ the \lq vacuum\rq~(absorbing state I) is the only
attractor of the model while the other two absorbing states
(II and III) are never reached from disordered initial states.
Consequently the \lq vacuum\rq~appears, in effect, to be the
unique absorbing state in the subcritical region. The dynamic
critical exponents characterising the phase transition
in this model indicate that the transition belongs to the class
of DP, in agreement with the conjecture of~\cite{Janssen1981,
Grassberger1982}.

\indent On the other hand, the phase transition occuring in
model $A$ of~\cite{Grassberger1984} belongs to the PC class.
This model is yet another one-dimensional elementary probabilistic
cellular automaton defined by the evolution rule :

\begin{equation}
{t : \over {t+1 :}}~~~~~~{111 \over 0}~~\underbrace{{110 \over  }~~~~{011 \over  }}_{\begin{array}{lll}
                                                                                              0 & \mbox{with probability} & p\\
                                                                                              1 & \mbox{with probability} & 1 - p
                                                                                       \end{array}}~~{101 \over 0}~~~~{010 \over 1}~~~~{100 \over 1}~~~~{001 \over 1}~~~~{000 \over 0}
\label{eq:modelA}
\end{equation}

\vspace{0.5cm}

\noindent It is remarkable that the three absorbing states of this model
are exactly the same as those of model (\ref{eq:model}). However,
contrary to their nature in model (\ref{eq:model}), the mutually
symmetric absorbing states II and III occur as attractors of this
model for $p < p_c$ while aborbing state I (the \lq vacuum\rq) is
never reached from disordered initial states. It appears that
the contrast in the nature of the absorbing states between
the two models places them in different universality classes.
In that case the non-DP behaviour of model (\ref{eq:modelA}) must
be due to the degeneracy in the absorbing state in the subcritical
region, thus supporting the view of~\cite{Park1995, Hinrichsen1997}.

\section*{Acknowledgment}

\indent I am grateful to Professor Bikas K. Chakrabarti for discussions
and for critically reading the manuscript. The work was supported
by CSIR, Government of India.

\newpage

\section*{Figure Captions}

\vspace{1cm}

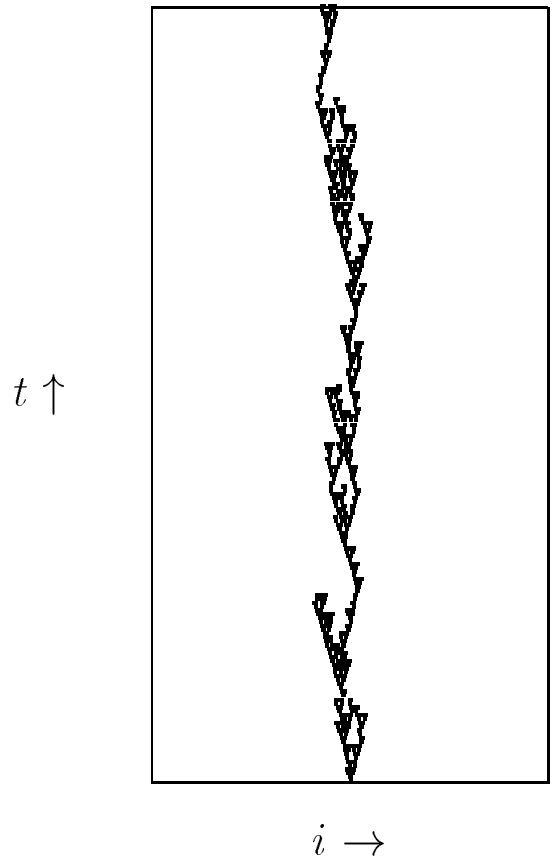
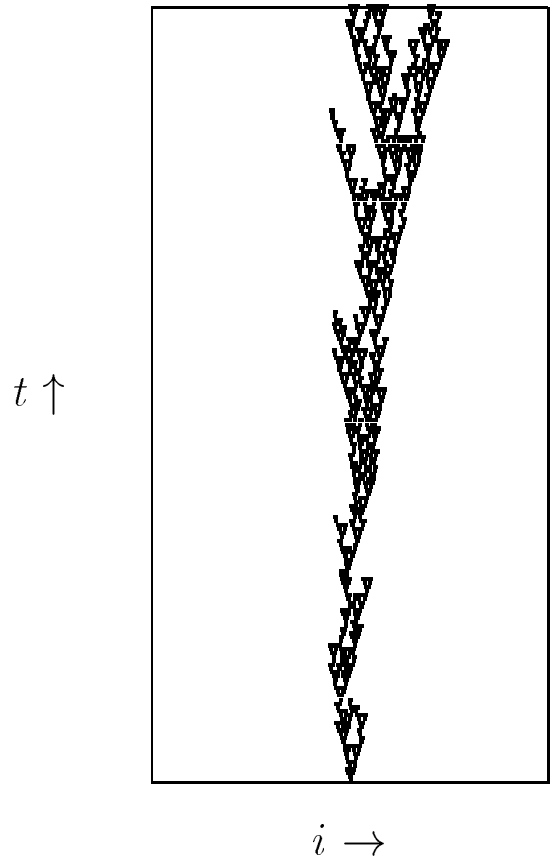
\noindent {\bf Figure 1.}~~Typical examples of evolution from a single
occupied site following rule (\ref{eq:model}) : (a) a case in the
subcritical region ($p = 0.74$) and (b) a case in the supercritical
region ($p = 0.76$). The first $200$ time-steps if the evolution process
are shown.

\vspace{1cm}

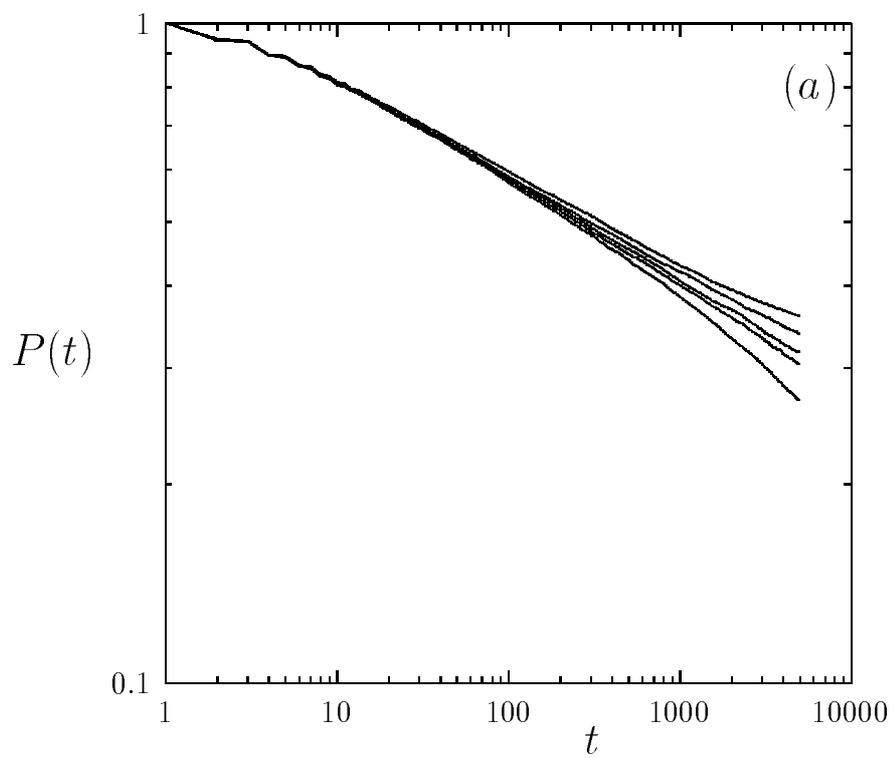
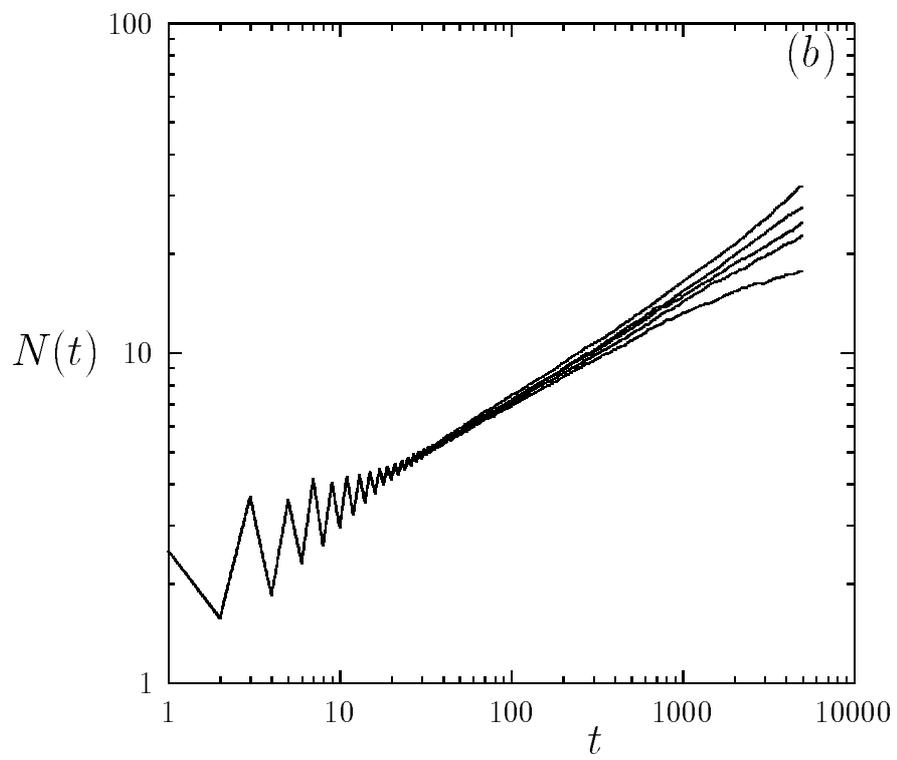
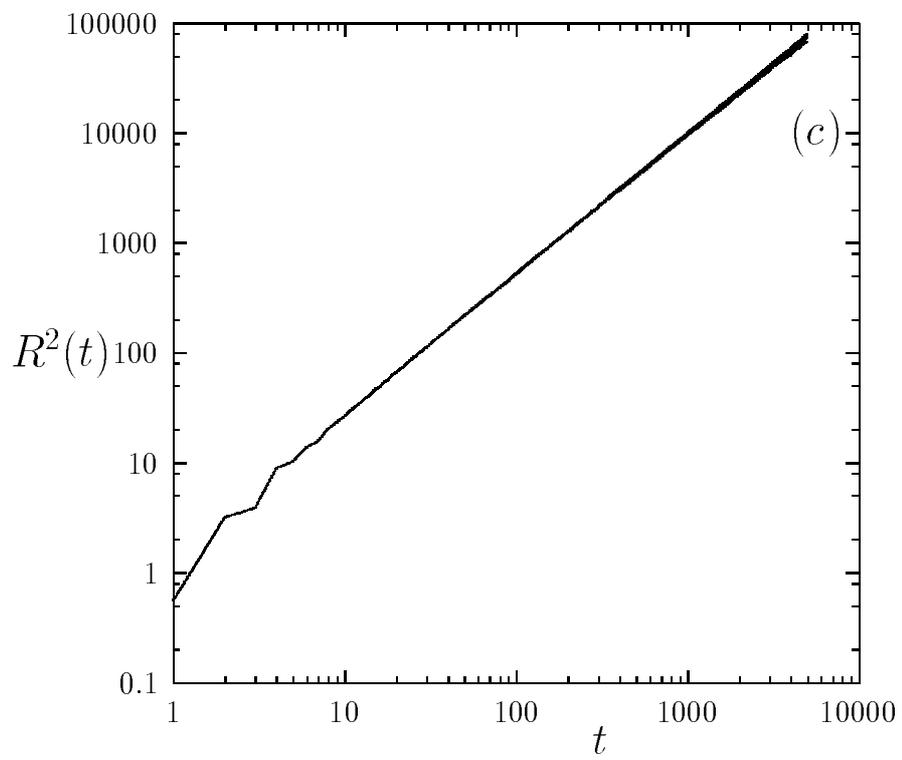
\noindent {\bf Figure 2.}~~Results for three dynamic properties
from monte-carlo simulations of the model : (a) survival probability,
(b) average number of occupied sites and (c) mean square radius
of the evolving cluster. Each of the three panels contains five
curves that correspond to $p$ = $0.753$ (top), $0.752$, $0.7515$,
$0.751$, and $0.75$ (bottom) respectively.

\vspace{1cm}

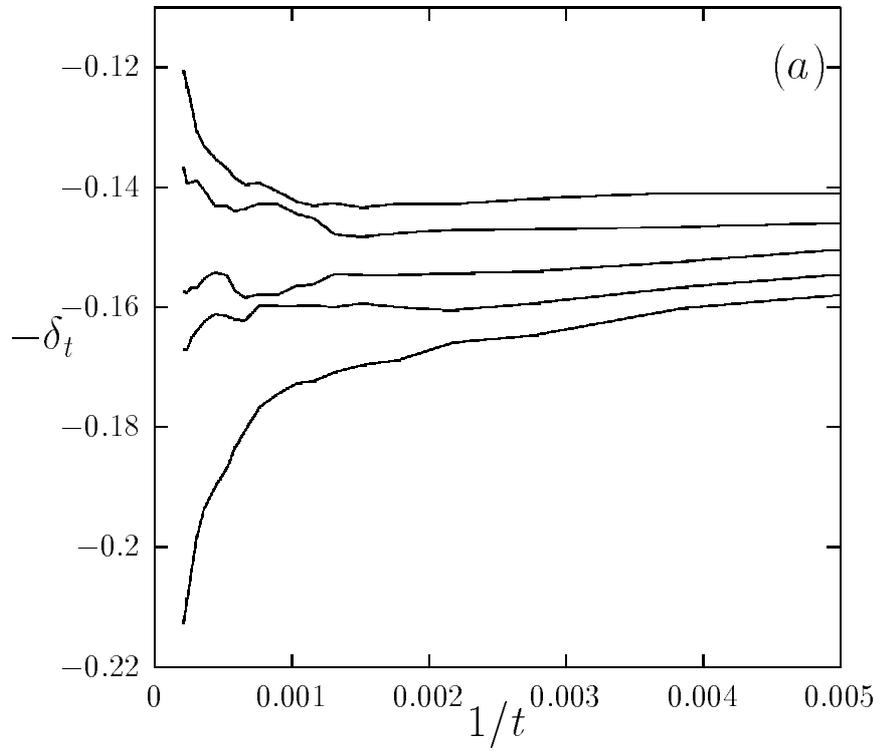
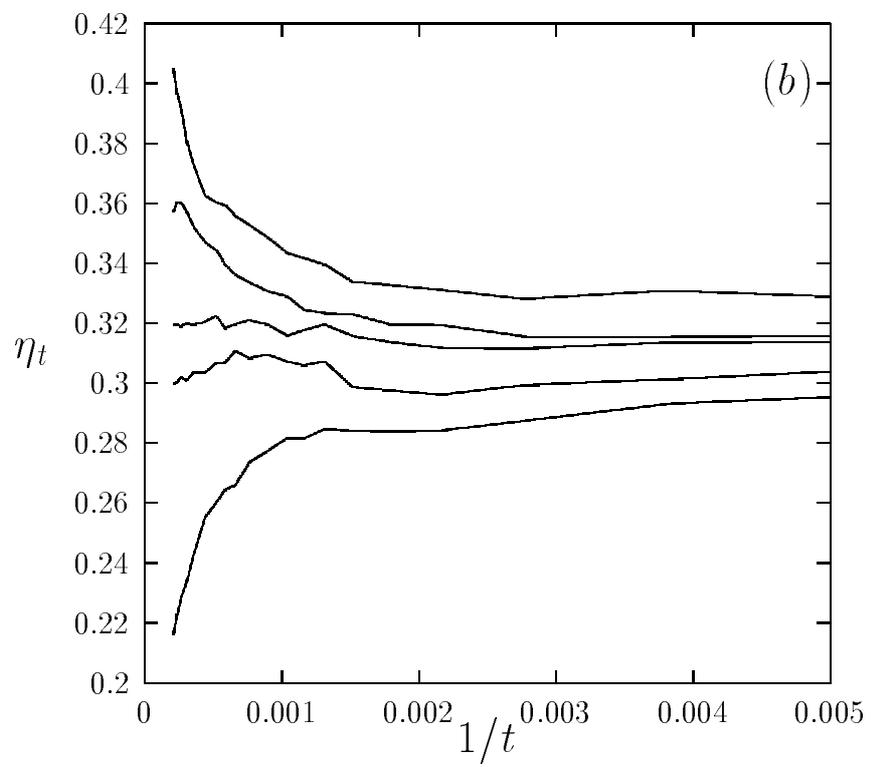
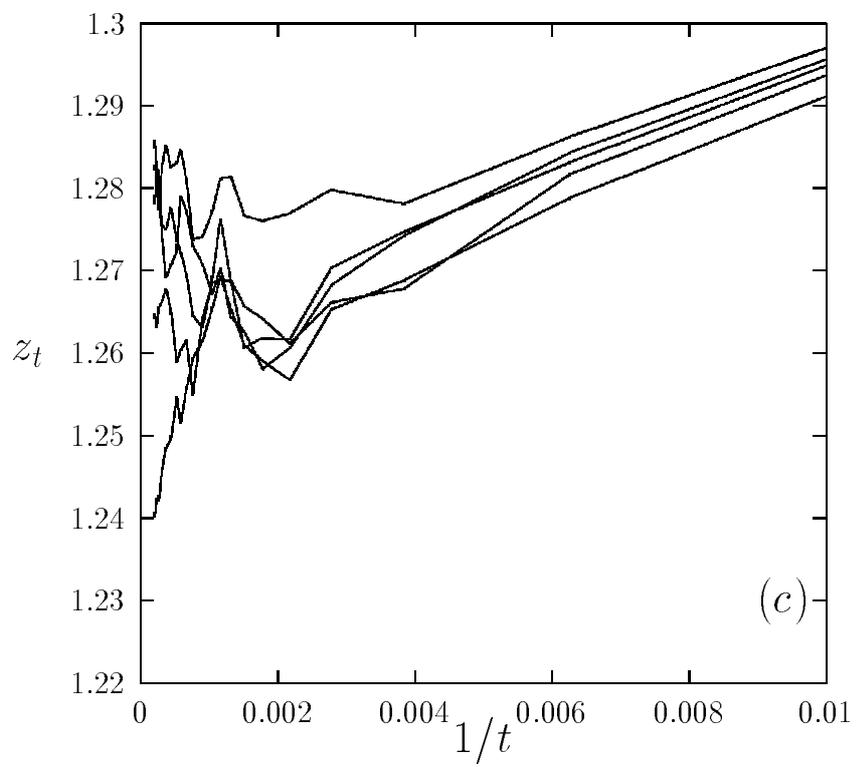
\noindent {\bf Figure 3.}~~The effective exponents measured as the
local slopes of the curves shown in figure 2. For large $t$, data
have been averaged over many time-steps in order to suppress
fluctuations.

\vspace{1cm}

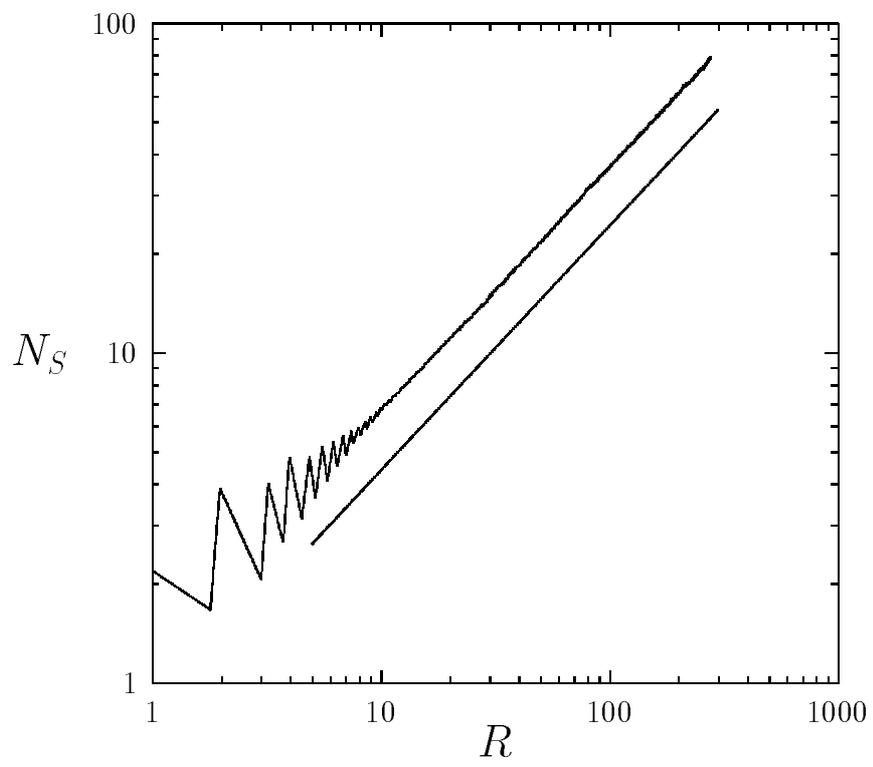
\noindent {\bf Figure 4.}~~The log-log plot of the average number
of occupied sites as a function of the root-mean-square radius of
the clusters at $p = p_c$. The slope of the curve gives the fractal
dimension of the clusters at fixed time. The straight line drawn
below the curve is the graph of $N_s = {\rm const.}~R^{0.74}$.

\end{document}